\documentclass{llncs}
\usepackage{graphicx}
\usepackage{multirow}
\usepackage{cleveref}
\usepackage{amsmath}
\usepackage{amssymb}
\usepackage{bbm}
\usepackage{algorithm}
\usepackage{algpseudocode}
\usepackage{todonotes}

\usepackage{enumitem}
\usepackage{breqn}
\usepackage{comment}

\usepackage[hyphens]{url}
\usepackage{hyperref} 
\usepackage{float}
\usepackage{dirtytalk} 
\usepackage{ntheorem}
\usepackage{hhline}
\newtheorem*{mytheorem}{Theorem}[section]

\DeclareMathOperator{\e}{{e}}

\newcommand{\chain}[1]{C^{#1}}
\newcommand{\block}[2]{B^{#1}_{#2}}

\newcommand{\activebranch}[1]{\textsf{active}(\chain{#1})}
\newcommand{\head}[1]{\textsf{head}(\chain{#1})}
\newcommand{\lab}[1]{\mathsf{label}(#1)}
\newcommand{\tx}[1]{t_{#1}}
\newcommand{\fulltx}[3]{{#1} \rightarrow {#2} : {#3}}
\newcommand{\txset}[1]{T_{#1}}
\newcommand{\acca}{a}
\newcommand{\accb}{b}
\newcommand{\acc}[1]{\acca_{#1}}
\newcommand{\acctree}{F}

\newcommand{\hash}[1]{H({#1})}
\newcommand{\height}[1]{h({#1})}
\newcommand{\acctreelag}{\Delta}

\begin{document}
\title{Hydra: A Multiple Blockchain Protocol for Improving Transaction Throughput}
\author{Rowel G\"{u}ndlach\inst{2}, Jaap-Henk Hoepman\inst{1}, Remco van der Hofstad\inst{2},\\Tommy Koens\inst{1}, Stijn Meijer\inst{1}}

\institute{Radboud University, Institute of Computing and Information Sciences,\\6525 EC Nijmegen, The Netherlands,\\
\email{jhh@cs.ru.nl, tkoens@cs.ru.nl, stijn.meijer@ing.com},
\and
Eindhoven University of Technology,
Department of Mathematics and Computer Science,
5600 MB Eindhoven, The Netherlands,\\
\email{r.c.gundlach@student.tue.nl, rhofstad@win.tue.nl}}

\maketitle

\begin{abstract}
Improving transaction throughput is one of the main challenges in decentralized payment systems.
Attempts to improve transaction throughput in cryptocurrencies are usually a trade-off between throughput and security or introduce a central component.

We propose Hydra, a decentralized protocol that improves transaction throughput without the security trade-off and has no central component.
Our novel approach distributes blocks over multiple blockchains.
Hydra makes a trade-off between transaction throughput and finality, the time it takes to stabilize the record of a transaction in the blockchain.
We rigorously analyze the double spend attack in a multiple-blockchain protocol.
Our analysis shows that the number of transactions per second can be increased significantly while finality is within acceptable boundaries.
\end{abstract}

\section{Introduction}
\label{intro}
Decentralized cryptocurrencies like Bitcoin \cite{nakamoto2008bitcoin} serve as a payment system without the need for a centralized trusted third party. However, there is a large gap between the transaction throughput of decentralized cryptocurrencies and that of mainstream payment processors. Bitcoin, for example, can currently process only a maximum of 7 transactions per second (tps) \cite{mccorry2016towards}, compared to VISA's 2000 tps \cite{croman2016scaling}. If decentralized cryptocurrencies are to challenge centralized payment processors, then their transaction throughput must clearly be improved.

Transaction throughput in blockchain-based cryptocurrencies can be improved by increasing the frequency in which sets of transactions (called blocks) are generated by network nodes.
Although increasing the block frequency in Bitcoin would increase the transaction throughput, it also leads to a critical instability where the system is in disagreement over its current state with regards to its most recent transactions \cite{eyal2016bitcoin}.
This instability is caused by forks, where multiple valid blocks are proposed that all point to the same previous block.

In this work we propose Hydra, a decentralized blockchain protocol.
Hydra is in principle similar to Bitcoin, but with an increased block frequency.
Hydra achieves this performance improvement by randomly distributing the blocks over multiple independent blockchains.
The trade-off Hydra makes is that of transaction finality, which slightly increases compared to Bitcoin.
Additionally, Hydra equally distributes mining capacity over multiple chains so that all chains grow at the same rate, on average.
Hydra achieves this by delaying the decision to which chain a block is appended.
To the best of our knowledge, this way of improving transaction throughput has not been examined before in the literature.

The remainder of our work is structured as follows: In Section~\ref{background} we we introduce the terms and concepts of blockchain technology.
We make the following three contributions.
First we describe Hydra in Section~\ref{onhydra}.
Our second contribution is a rigorous analysis of the effect of an increased block frequency on transaction finality in our multi-blockchain protocol in Section~\ref{detailedanalysis}.
As the instability of the system caused by forks is critical for any blockchain, we solely analyze the effect of an increased block frequency in our work.
Our third contribution is a discussion of the results of our analyses in Section~\ref{results}.
We determine the probability of a successful double spend in Hydra, and compare the outcomes to Bitcoin.
Here we show that in Hydra it takes slightly longer to reach a small probability of a fork occurring.
Lastly, we describe the work related to improving transaction scalability in Section~\ref{relatedwork}.
In Section~\ref{limitations} we discuss the limitations of Hydra and discuss its future work, and in Section~\ref{conclusion} we conclude our paper.

\section{Background}
\label{background}
Blockchain implements an immutable ledger, which is stored on every participating node in the network and contains all the valid transactions that ever took place.
Ledger accounts are a representation of a cryptographic public key.
Ownership of the account value can be transferred by a transaction.
Bitcoin transactions are based on an unspent transaction output (UTXO) model, which works as follows.
Each transaction consists of inputs and outputs.
Each input is cryptographically signed to prove ownership of the value transferred by a previous transaction and each input contains a reference to a previous transaction's output.
An output contains the requirements that are set for proving ownership.
When the cryptographic signature of the input matches with the requirements set in the output, the sender has proven ownership.
Each transaction consumes an output and therefore each output can be used only once.

Transactions are sent to all nodes in the network.
To achieve unique transaction ordering, network participants called miners bundle transactions in a set called a block.
Note that each block contains a sequential number called the block height.
Each block consists of a block body and block header.
The body contains a set of valid transactions chosen by the node.
From the body transactions a Merkle tree (a type of binary hash tree) is calculated and its root is added to the header.
The header also contains the hash of its predecessor block and a nonce.
Each miner calculates a hash over the header, with the aim of finding a specific hash value.
This value must be smaller than the current target.
This target is shared amongst all miners and determines how hard it is to find the hash value.
A miner must sequentially increase the nonce and re-calculate the hash over the header to find a value that matches the target.
This process is called proof-of-work (PoW).
Once a valid hash has been found the block is broadcast to, verified and stored by all nodes.

Occasionally, nodes mine blocks almost simultaneously.
These blocks may contain different transactions, but point to the same previous block.
This results in two branches of the blockchain which is called a fork.
Forks are an issue because they represent possibly conflicting histories.
Forks are resolved due to the {\lq}longest chain{\rq} rule.
This rule states that only the chain with the most cumulative PoW is valid.
In essence, a fork divides the total computational power of all honest miners.
This allows an attacker to create blocks faster than the rest of the network, because the attacker's computational power is increased relative to the rest of the network.
This allows the attacker to perform a double-spend attack.
This attack can be described as follows.
Consider a transaction $t$ of the attacker that is present in the main chain $c$.
Also assume that the attacker received goods related to this transaction.
The attacker now creates a longer chain $c'$ that does not include transaction $t$.
Once the attacker publishes chain $c'$, transaction $t$ will be discarded.
In this attack scenario the attacker thus received goods for free.

\section{The Hydra Protocol}
\label{onhydra}
Hydra is designed to increase transaction throughput.
Our main idea is to use $N$ blockchains, where $N \geq 1$.
We distribute blocks randomly over chains, which allows to significantly increase the block frequency.
Despite the high block frequency, the probability of a fork is divided evenly over $N$ different chains.

\subsection{Hydra Accounts and Transactions}
Hydra is UTXO based, similar to Bitcoin.
To abstract from the UTXO concept, we consider that an account $a$ stores some monetary value $v$.
Account numbers are the hash of a public key.
The corresponding private key controls the account, by signing transactions that transfer value from this account.
A transaction $\tx{}$ transfers value from one single account to another single account.

We write $\fulltx{\acca}{\accb}{v}$ to denote that a transaction transfers $v$ units of value from account $\acca$ to account $\accb$.
Each transaction $\tx{}$ is assigned a label $\lab{\tx{}} \in \{0,\ldots,N-1\}$ that determines on which chain this transaction must be recorded.
The label of a transaction is based on its incoming account $\acca$: $\lab{\fulltx{\acca}{\accb}{v}}=\acca \bmod N$. 
This ensures that all transactions from the same account are always recorded in the same chain.
This allows us to determine the validity of a transaction without considering the state of the other chains.

\subsection{The Account Tree}
\label{accounttree}
Hydra does not only maintain $N$ blockchains, but also an account tree $F$ which records a value $F[a]$ for each account $a$.
The account tree helps to reduce space requirements and to speed up transaction validation.
Each node maintains a copy of the account tree, a data structure that can be used to store all bindings, e.g. a public key and an account value.
The account tree $\acctree$ is associated with a certain height $h()$.
Initially, $\height\acctree=0$, and all accounts are empty.
An account tree corresponds to the state of the accounts \emph{after} processing all transactions in the blocks up to and including height $\height{\acctree}$ on the valid branches of all chains $\chain{i}$.
Hydra defines a constant $\acctreelag$ that defines the {\lq}lag{\rq} of the account tree w.r.t. the height of the blockchain $\height{\chain{i}}$. The account tree moves forward along the blockchains when all chains have grown high enough to do so. To be precise, we always ensure that $\height{\acctree}$ is $\acctreelag$ less than the shortest chain.

\subsection{Hydra Chains}
\begin{figure}[!ht]
\centering
\includegraphics[width=4.5in]{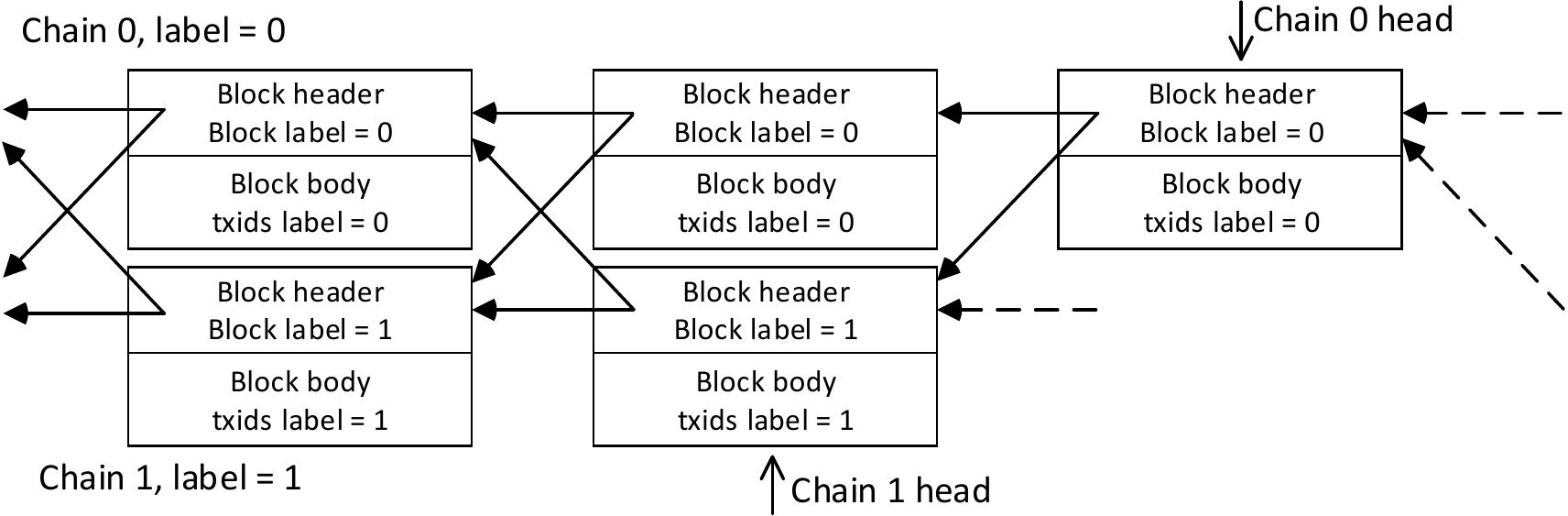}
\caption{Two simplified Hydra blockchains, containing two heads.}
\label{hydrabcs}
\end{figure}

Hydra uses $N$ blockchains.
Each chain is assigned a fixed label $\lab{C^{i}} \in \{0,\ldots,N-1\}$, see Figure~\ref{hydrabcs}.
The head $\head{i}$ of a chain $\chain{i}$ is the block with the largest height in the chain.
We call the path from the head to the genesis block on (the first block of a chain) chain $\chain{i}$ the \emph{active branch} and write $\activebranch{i}$.
As in Bitcoin, this active branch is considered to contain the transactions that are {\lq}active{\rq} and contribute to the current state.
We define the height of a chain $\height{\chain{i}}$ to be the height of its head, i.e. the length of its active branch.

\subsection{Mining Hydra Blocks}
Like transactions, blocks $\block{}{}$ have a label $\label{\block{}{}}$ that determines to which chain a block must be added.
Blocks obtain their label only \emph{after} they are successfully mined.
In fact, the label of a block equals its hash modulo $N$.
This, in essence, distributes the network's total mining capacity over multiple chains.
However, this raises the question of which transactions to add to a block, as the transactions have fixed labels while the block only receives its label after it is mined.
The solution is to prepare sets of transactions, one for each label and only include the set that corresponds to the final label of the block.

The details are as follows. A miner that wishes to mine a block constructs $N$ sets of valid transactions $\txset{0},\ldots,\txset{N-1}$.
Thus $\txset{i}$ contains only transactions with label $i$. It then computes the Merkle root hash $\hash{\txset{i}}$ for each $T_{i}$ and stores these in the block header, together with the hashes of the head of each of the chains $\chain{0},\ldots,\chain{N-1}$.
Note that from each of these sets of hashes a Merkle root could be created to save space in the block header.
However, the space saved has no significant impact on Hydra's transaction throughput.

When the miner finds a matching nonce, such that the hash of the block header becomes less than the target value, the block label (say it is $i$) is fixed.
The miner then moves the transactions in $\txset{i}$ to the block body, and stores the transactions in all other sets $\txset{j}$.
These will be considered again for inclusion in another block.
The block with label $i$ is only added to chain $\chain{i}$, to which it already links as it includes the head of that chain.
Note that a miner could output blocks only with a particular label, but at the cost of reducing the miner's block rate by $N$.

\subsection{The Effect of a Single Fork on Multiple Chains}
As in Bitcoin, transactions that were active first suddenly may become inactive due to a fork.
Because Hydra maintains multiple chains, such a change in active transactions on one chain could have an impact on the validity of other chains.
Consider the following example, where we have two chains $\chain{0}$ and $\chain{1}$, and four accounts $a$, $b$, $c$, and $d$.
Let the current state of the account tree hold the following values for the accounts: $\acctree[a]=10$, $\acctree[b]=0$, $\acctree[c]=20$, and $\acctree[d]=0$.
Now consider a transaction $\fulltx{a}{c}{5}$ added to a block at height $h$ on chain $0$.
The value in account $c$ now equals $25$.
This is followed by a transaction $\fulltx{c}{b}{25}$ which is added to a block at height $h+1$ in chain $1$.
Now suppose a fork in chain $0$ occurs, and a new active branch is created with a transaction $\fulltx{a}{d}{10}$ added to the block at height $h+1$ on chain $0$.
The transaction $\fulltx{a}{c}{5}$ on chain $0$ is no longer active, but this \emph{invalidates} the transaction $\fulltx{c}{b}{25}$ on chain $1$.

This scenario shows that there is a risk of cascading invalidations due to a single chain fork.
This increases the adversary's power beyond the single chain, thus reducing security.
It also wastes mining power.

\subsection{Block Processing and Validity of Transactions}
\begin{figure}[!ht]
\centering
\includegraphics[width=4.8in]{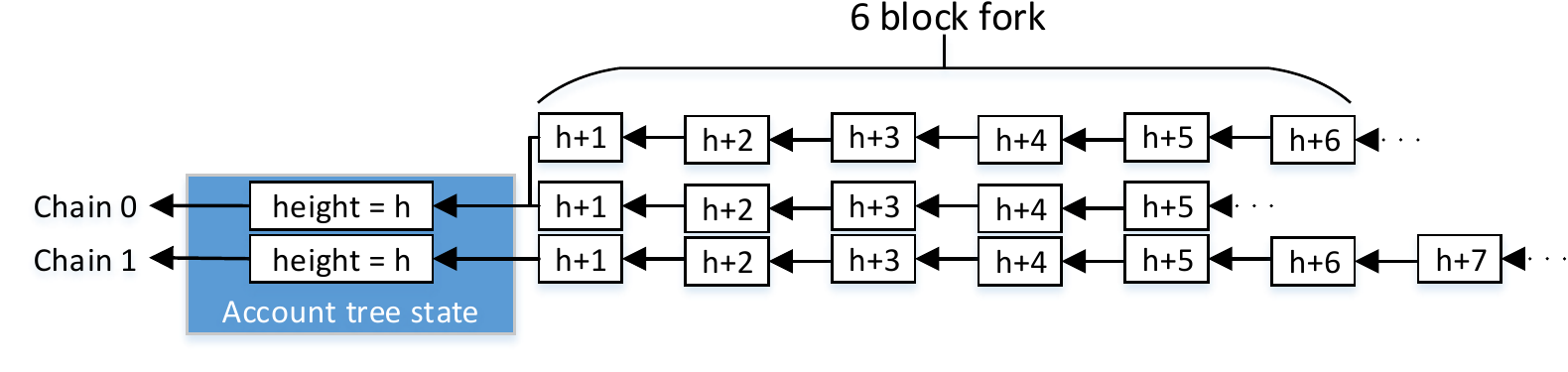}
\caption{Severing the link between a fork and account state ($z = 5$).}
\label{hydrasolution}
\end{figure}

We sever the effect of a fork and multiple chains by waiting at least $\acctreelag$ blocks on all chains.
Consider a state of accounts, which is set by the transactions in block height $h-1$ of all $N$ chains.
A new state of accounts is computed from the transactions at block height $h$, of all $N$ chains. This occurs when the shortest of these chains contains at least $\acctreelag$ new blocks from height $h$, see Figure~\ref{hydrasolution}.
Now, if a fork at height $h+1$ occurs on a particular chain, then only that single chain is affected by the fork because the account tree is not updated yet.

However, since there is a brief delay in transaction processing, an account can be charged more than it holds.
Note that in the above scenario the transaction $\fulltx{c}{b}{25}$ is considered valid because there was a previous transaction $\fulltx{a}{c}{5}$ that increased the current value of account $c$ beyond what was recorded in the account tree.
We therefore introduce a {\lq}strict validity{\rq} rule to prevent this.
If we restrict transactions in the current active branch of a chain to never spend more (in total) than what is recorded in the account state for that account, then the scenario is prevented.
Here we use the fact that transactions from a particular account always have the same label and hence end up on the same chain.

Consider the account tree $\acctree$ at height $\height{\acctree}$. Consider a transaction $\fulltx{\acca}{\accb}{v}$. Let $i=\lab{\acca}$ be $\acca$'s label. Consider the active branch of chain $\chain{i}$, call it $X^i$, on which this transaction will end up. Consider all blocks $\block{i}{h+1},\ldots$ with height larger than $h$ on $X^i$.
There are at least $\acctreelag$ of these.
Consider all transactions from account $\acca$, i.e. $\fulltx{a}{\acc{i}}{v_i}$, in all these blocks.
Let $S(\acca)=\sum_i v_i$ be the sum of all values transferred from account $\acca$ on the active branch that are not accounted for yet in the value $\acctree[\acca]$ stored in the account tree for account $\acca$.
Transaction $\fulltx{\acca}{\accb}{v}$ is considered valid if $\acctree[\acca]-S(a) > v$.

To advance the account tree from height $h$ to height $h+1$, for each chain $C^{i}$ all transactions in the block on height $h+1$ are collected and executed to update the values of all the accounts.
Because of the \lq strict validity{\rq} rule the order in which transactions are processed to update the account tree is no longer relevant.

\section{Detailed Double Spend Analysis}
\label{detailedanalysis}
In this section we model the double spend attack in a multi-blockchain protocol, and find the equation that allows for determining the probability of a successful double spend in Hydra. 
We choose the following values as variables:
\begin{enumerate}
\item $\mathbb{P}[A]$, as the probability $\mathbb{P}$ of a successful double spend on any chain.
\item $p$ and $q$, representing the mining capacity of the honest users and attacker respectively, where $q+p=1$. We assume $p>q$.
\item $N$, representing the number of chains of the blockchain protocol.
\item $w$, the tipping point, the number of blocks that have to be present on each chain before a take-over can occur.
\item $\lambda$ and $\mu$, describing the block's arrival intensity per chain for the honest nodes and attacker respectively.
These values depend on the relative mining capacity, the number of chains and the expected time between finding blocks (predetermined, $T_0$). This leaves us with $\lambda= \frac{p}{T_0\cdot N}$ and $\mu= \frac{q}{T_0\cdot N}$.
\end{enumerate}
We start to set up the base of our formula.
It is important to note that a take-over can occur on each of the $N$ chains.
We can, however, make a distinction between these chains, namely that there always is exactly one short chain $S$.
This chain receives its $w$'th block last and will surely have $w$ blocks when the tipping point occurs.
The $N-1$ other chains are denoted by $C$. 
We are sure that there are at least $w$ blocks on these chains after the tipping point.

It is important to note that all chains behave independently.
Based on the work by Bowden et al. \cite{bowden2018block}, we assume that the time between finding blocks is exponentially distributed and the blocks arrive according to a Poisson process.

Upon arrival they are assigned to chain $j$ with probability
$1/N$, for all $ j \in \lbrace 0,...,N-1 \rbrace$.
As this procedure is independent from past arrivals, we find that on every chain a separate independent Poisson process occurs.
The Poisson process on a single chain is independent of the Poisson processes of other chains.

We present the base equation for the probability of a successful double spend attack $\mathbb{P}[A]$ in \eqref{eq:F1}.
To determine this probability we have to find three distinct formulas.
Part one of the equation is the probability of a take-over on the short chain $\mathbb{P}[A_{S}]$, given that it took $X$ time to get to the tipping point.
Part 2 describes the probability of a take-over on any of the other chains, $\mathbb{P}[A_{C}]$.
Finally, in part three we determine the density function for the time to the tipping point.

\begin{equation}\label{eq:F1}
\begin{split}
&\mathbb{P}[A] = 1-\int\limits_{x=0}^{\infty}\underbrace{(1-\mathbb{P}[ A_S \mid X=x])}_{\text{part 1}}\cdot\underbrace{(1-\mathbb{P}[A_C \mid X=x])^{N-1}}_{\text{part 2}} \cdot \underbrace{f_X(x)\mathrm{d}x.}_{\text{part 3}}\\
\end{split}
\end{equation}

\subsubsection{Part 1: take-over on the short chain.}
On the short chain there are only two possible cases.
The attacker either has already taken over the short chain by finding more than $w$ blocks after the tipping point, or he is not ahead of the honest nodes after $x$ time units.
In the first case the probability of a take-over is 1.
In the latter case, the probability of a take-over simplifies to a one-dimensional random walk.
Denote the number of blocks found by the attacker with $b$, then we consider the case $b \leq w$, such that the honest nodes have a non-negative head start of $w-b$ blocks.
If the first block after the tipping point is found by the honest nodes or attacker, this difference $w-b$ increases or decreases by one respectively.
When this difference ever becomes negative, a take-over has occurred.
We can model this as a random walk $S_n$ that has a non-negative starting point $S_0=w-b$, and we are interested in the probability of the random walk ever reaching $-1$.
The probability to go one level up or down is the probability that the next block found is for the honest nodes or the attacker, and are given by $\frac{p}{q+p}$ and $\frac{q}{q+p}$ respectively.
These are independent of past events due to the memoryless property of the exponential distribution that models the arrival times of blocks for attacker and honest nodes. 
The probability of ever reaching $-1$ is given by  
\begin{equation}\label{eq:RW}
\mathbb{\exists}\:{n} \in \mathbb{N} \text{ such that } {P} [S_n=-1 \mid S_0=w-b]=\text{min}\left\{\left(\frac{q}{p}\right)^{w-b+1},1\right\}.
\end{equation}
As we assumed the capacity of the attacker to be smaller than that of the honest nodes, we find the probability of an eventual take-over is denoted by $ \left(\frac{q}{1-q}\right)^{w-b+1}$.

We assumed the time it took for the honest nodes to reach $w$ to be fixed at $x$ time units, so we can calculate the probability that the attacker found $i$ blocks in this time by conditioning on this number $B_A$.
We find that the probability of a successful take-over on the short chain $S$ is
\begin{equation}\label{eq:F2}
\begin{split}
 \mathbb{P}[A_S \mid X=x]=&\sum\limits^{\infty}_{i=0}\mathbb{P}[A_S \mid X=x,B_A=i]\cdot\mathbb{P}[B_A=i]\\
 =&\sum\limits^{\infty}_{i=0} \left(\mathbbm{1}[i\leq w]\cdot\left(\dfrac{q}{1-q}\right)^{w-i+1}+\mathbbm{1}[i>w]\right)\cdot \e^{-x\mu}\cdot\dfrac{(x\mu)^i}{i!}.
\end{split}
\end{equation}

\subsubsection{Part 2: take-over on the other chains.}
For the remaining chains $C$, the number of blocks obtained by the honest nodes is not fixed anymore, since blocks are randomly distributed over chains.
Instead of conditioning on the number of blocks found by the attacker immediately, we condition over the difference of blocks ($Z$) first.
In this case \lq difference{\rq} can either be zero, or positive (a head start for the honest nodes), or negative (a head start for the attacker).
Again, in the latter case the probability of a take-over is 1.
In the former case the probability can be modeled as a random walk, and we can apply (\ref{eq:RW}).

We split the events of a positive and a negative head start in \eqref{eq:F3}.
For a non-negative head start of $z$ we condition over all possibilities where the difference of blocks between the honest nodes and the attacker is exactly $z$.
This can be done by, again, conditioning on the number of blocks for the attacker.
Here it is important to note that the honest nodes will always have at least $w$ blocks, otherwise the tipping point would not have occurred. 

\begin{equation}\label{eq:F3}
\begin{split}
 \mathbb{P}&[A_C \mid X=x]\sum\limits^{\infty}_{z=-\infty}\mathbb{P}[A_C \mid X=x,Z=z]\cdot\mathbb{P}[Z=z]\\
=&\sum\limits^{\infty}_{z=0}{\bigg[}{\bigg(}\dfrac{q}{1-q}\bigg)^{z+1}\cdot\bigg(\sum\limits_{n=\text{max}\{w,z\}}^{\infty}\e^{-x\mu}\dfrac{(x\mu)^{n-z}}{(n-z)!}\e^{-x\lambda}\dfrac{\dfrac{(x\lambda)^{n}}{n!}}{1-\sum\limits_{k=0}^{w-1}\e^{-x\lambda}\dfrac{(\lambda x)^k}{k!}}\bigg)\bigg]\\
+&\sum\limits^{\infty}_{z=1}\bigg[1\cdot\bigg(\sum\limits^{\infty}_{n=w}\e^{-x\mu}\dfrac{(x\mu)^{n+z}}{(n+z)!}\e^{-x\lambda}\dfrac{\dfrac{(x\lambda)^{n}}{n!}}{1-\sum\limits_{k=0}^{w-1}\e^{-x\lambda}\dfrac{(\lambda x)^k}{k!}}\bigg)\bigg]
\end{split}
\end{equation}

\subsubsection{Part 3: Density of the time to the tipping point.}
To find the density of the time to the tipping point, we first look at the distribution function.
Note that when we set $X_i$, the time it takes for chain $i$ to have at least $w$ blocks, where $i \in \lbrace 0,...,N-1\rbrace$, the time to the tipping point $X$ is equal to the maximum of the set of $X_i$'s.
Only when all chains have at least $w$ blocks, the tipping point occurs.
Due to our assumption that the times between finding blocks is exponential, we find that all $X_i$ are defined as a sum of exponentially distributed stochastic variables.
As a result we find that the distribution function $F_{X_i}(x)$ of a single $X_i$ equals
\begin{equation}\label{eq:F4}
   F_{X_i}(x)= 1-\e^{-\lambda x} \sum\limits^{w-1}_{k=0}\dfrac{(\lambda x)^k}{k!}
\end{equation}
\noindent
and due to all the $X_i$ being independent and identically distributed from each other, we find that the distribution function $F_X$ of $X = \smash{\displaystyle\max_{i}^{N}}\:X_i$ is given by
\begin{equation}\label{eq:F5}
   F_X(x)= \left[1-\e^{-\lambda x} \sum\limits^{w-1}_{k=0}\dfrac{(\lambda x)^k}{k!}\right]^{N}.
\end{equation}
\noindent
Now, we differentiate the above defined distribution function to find the density function, to obtain
\begin{equation}\label{eq:F6}
          f_X(x)=N\cdot\left[\lambda\e^{-\lambda x}\left( \dfrac{(\lambda x)^{w-1}}{(w-1)!} \right)\right]\cdot\left[1-\e^{-\lambda x} \sum\limits^{w-1}_{k=0}\dfrac{(\lambda x)^k}{k!}\right]^{N-1}.
\end{equation}

\subsubsection{The final equation.}
With these three different formulas, we can construct the formula that gives the probability of a successful take-over event on any of the $N$ chains.
The final equation consists of (\ref{eq:F1}), where we include (\ref{eq:F2}), (\ref{eq:F3}) and (\ref{eq:F6}).
Below we present the simplified version of these substitutions, which allows for implementing in a software program.

\begin{mytheorem}[The final equation]
Let $H$ be a Hydra system with $N$ chains and q be the capacity of the attacker, where q$<$0.5 and $w$ denotes the number of blocks to the tipping point. Then the probability of a successful take-over is given by 
\end{mytheorem}

\begingroup\makeatletter\def\f@size{8}\check@mathfonts
\def\maketag@@@#1{\hbox{\m@th\large\normalfont#1}}%
\begin{equation}\label{eq:F7}
\begin{split}
A(q,N,w)=&1-\int\limits_{x=0}^{\infty}
\left(\e^{-x\mu}\sum\limits^{w}_{i=0}\left(1-\left(\dfrac{q}{1-q}\right)^{w-i+1}\right)\dfrac{(x\mu)^i}{i!} \right)\cdot
\Bigg(\left[1-\e^{-\lambda x} \sum\limits^{w-1}_{k=0}\dfrac{(\lambda x)^k}{k!}\right]\\
-&\Bigg(\sum\limits^{\infty}_{z=0}\left[\left(\dfrac{q}{1-q}\right)^{z+1}\cdot\left(\sum\limits_{n=\text{max}\{w,z\}}^{\infty}\e^{-x\mu}\dfrac{(x\mu)^{n-z}}{(n-z)!}\e^{-x\lambda}\dfrac{(x\lambda)^{n}}{n!}\right)\right]\\
+&\sum\limits^{\infty}_{z=1}\left[1\cdot\sum\limits^{\infty}_{n=w}\e^{-x\mu}\dfrac{(x\mu)^{n+z}}{(n+z)!}\e^{-x\lambda}\dfrac{(x\lambda)^{n}}{n!}\right]\Bigg)\Bigg)^{N-1}
\cdot N\left[\lambda\e^{-\lambda x}\left( \dfrac{(\lambda x)^{w-1}}{(w-1)!} \right)\right] \mathrm{d}x.
\end{split}
\end{equation}
\noindent

\section{Probability of a Successful Double Spend}
\label{results}
In this section we compare the results of the analytic equation (\ref{eq:F7}) and the simulation results on Bitcoin provided by Rosenfeld \cite{rosenfeld2014analysis}, in Table~\ref{tab:comparison}.
We used numerical integration methods\footnote{\url{implemented by Wolfram Mathematica version 10,https://www.wolfram.com/mathematica/}} to generate the results.

\begin{table}
    \centering
     \caption{Comparing Bitcoin \cite{rosenfeld2014analysis} and Hydra ((\ref{eq:F7}), $N=32$). The probability of a successful take-over.}
     \setlength{\tabcolsep}{3pt}
    \begin{tabular}{|c|c|c|c|c|c|c|c|c|c|c|}
    \hline
    \multicolumn{2}{|c|}{$\downarrow$Results of }&\multicolumn{9}{c|}{$w$}\\
    \cline{3-11}
                \multicolumn{2}{|r|}{$ q\downarrow $}&1&2&3&4&5&6&7&8&9\\
               \hhline{|=|=|=|=|=|=|=|=|=|=|=|}
        \cite{rosenfeld2014analysis} & 6&0.1200& 0.0200& 0.0039& 0.0007& 0.0001& 0.0000& 0.0000& 0.0000& 0.0000\\
        \cline{1-1}\cline{3-11}
        \cline{1-1}\cline{3-11}
        (\ref{eq:F7}) &&0.1589&0.3235&0.0068&0.0010&0.0003&0.0000&0.0000&0.0000&0.0000
        \\
        \hhline{|=|=|=|=|=|=|=|=|=|=|=|}
        \cite{rosenfeld2014analysis}   &  16&0.3200& 0.1372& 0.0635& 0.0305& 0.0149& 0.0074& 0.0037& 0.0019& 0.0009\\
        \cline{1-1}\cline{3-11}
        \cline{1-1}\cline{3-11}
        (\ref{eq:F7})  && 0.7726&0.4823&0.2720&0.1472&0.0785&0.0416&0.0221&0.0117&0.0062\\
        \hhline{|=|=|=|=|=|=|=|=|=|=|=|}
        \cite{rosenfeld2014analysis}   &  26&0.5200& 0.3353& 0.2286& 0.1603& 0.1142& 0.0823& 0.0598& 0.0438& 0.0322\\
        \cline{1-1}\cline{3-11}
        \cline{1-1}\cline{3-11}
        (\ref{eq:F7})   &&0.9935&0.9621&0.8938&0.7943&0.6790&0.5627&0.4555&0.3626&0.2850\\
        \hhline{|=|=|=|=|=|=|=|=|=|=|=|}
        \cite{rosenfeld2014analysis}   &  46&0.9200& 0.8802& 0.8506&  0.8261&  0.8048&  0.7857&  0.7683&  0.7523&  0.7374\\
        \cline{1-1}\cline{3-11}
        \cline{1-1}\cline{3-11}
        (\ref{eq:F7})  &&1.0000& 1.0000& 1.0000&  1.0000&  1.0000&  1.0000&  1.0000&  1.0000&  1.0000\\
        \hline
    \end{tabular}
    \label{tab:comparison}
\end{table}
\noindent

We observe three significant differences in comparing the two outcomes.
First, the probability of a successful double spend in Hydra is significant higher than in Bitcoin when there are only a few blocks.
The reason for this is that Hydra blocks are randomly assigned to chains and each time a new block is created there exists a possibility of extending a chain which was already extended.
Second, the probability of a successful double spend in Hydra reaches 0, although this takes more time than in Bitcoin.
This can be explained by the fact that Bitcoin only holds one chain on which a double spend can be performed.
In contrast, Hydra contains $N$ chains, which allows for the possibility of $N$ attempts of a double spend.
Third, Hydra is slightly less resilient to a successful double spend attack when there is an imbalanced mining power distribution.

\section{Related Work}
\label{relatedwork}
This section describes previous work related to improving transaction scalability.

Aspen also uses multiple blockchains with the aim of scaling services \cite{gencer2016service}. In contrast, Hydra aims at scaling transaction throughput.

A trade-off between transaction throughput and centralization is made in HoneyBadger \cite{miller2016honey} and Omniledger \cite{kokoris2018omniledger} by using a Byzantine Fault Tolerant (BFT) protocol.
As Hydra uses PoW, Hydra can be fully decentralized.

In Bitcoin-NG \cite{eyal2016bitcoin} the block interval is used to create several mini-blocks by a single miner.
This, however, leaves the option for an attacker to perform a Denial of Service (DoS) attack on the current leader.
There also exist hybrid protocols, employing both PoW and a type of BFT protocol.
Examples include, ByzCoin \cite{kogias2016enhancing}, SCP \cite{luu2015scp}, and Elastico \cite{luu2016secure}.
In Hybrid protocols PoW is employed to form groups of nodes called shards. These shards then reach consensus by employing a BFT protocol.
Some of these protocols are vulnerable to DoS attacks.
In Hydra there exists no single leader, which makes a leader DoS not possible.

Finally, payment channels such as the Lightning Network \cite{poon2015bitcoin} and the Raiden Network \cite{raiden} significantly increase transaction throughput.
Some nodes may lock a large amount of funds which are required to sustain multiple channels. 
Therefore, a possible downside of payment channels is that it may lead to more centralization.
In contrast, Hydra is as decentralized as Bitcoin.

\section{Limitations and Future Work}
\label{limitations}
A quick back-of-the-envelope calculation shows that Hydra improves transaction throughput.
If we assume that Hydra consists of 32 chains, each transaction is 240 bytes, the block size is 1 MB, and the block frequency is set to 18 seconds, then Hydra would achieve (1 MB / 240 b) / 18 $\approx$ 231 tps, a significant improvement over Bitcoin's 7 tps. All chains are updated in approximately 32*18 $\approx$ 10 minutes, similarly to Bitcoin.
However, bandwidth, network latency, and the random assignment of blocks will influence the number of Hydra forks. An implementation of Hydra will show how the protocol will hold up in practice.

Furthermore, we have shown that Hydra is resilient to a double spend attack.
This invites for further analysis such as selfish mining \cite{eyal2014majority} and chain quality \cite{garay2015bitcoin}.

Finally, an attacker could attack a single chain by sending transactions from multiple chains to one chain. Here, a single chain is flooded with transactions, delaying other transactions.
This requires the attacker to spend a large amount on transaction fees. The exact impact of chain-flooding requires further analysis.

\section{Conclusion}
\label{conclusion}
In this study we have proposed Hydra, a protocol that improves transaction scalability within a single decentralized cryptocurrency.
Current solutions make a trade-off between transaction throughput and either security, or decentralization, or significantly deviate from the Bitcoin protocol.
Hydra's trade-off is between transaction throughput and finality.
In Hydra, the block frequency is significantly increased thus being able to process more transactions per second.
Our protocol shows that it is possible to distribute the probability of a fork over multiple chains.
Also, our labeling strategy safely shards mining power over these chains.
Although Hydra requires further analysis, in our work we have shown that parallel blockchains offer a promising scalability solution for decentralized cryptocurrencies. 

\subsubsection{Acknowledgement.}
RvdH was supported by the Netherlands Organisation for Scientific Research
(NWO) through Gravitation-grant NETWORKS-024.002.003 and VICI-grant
639.033.806.


\end{document}